\documentclass[manuscript]{aastex}
\usepackage{rotating}

\newcommand{\Teff}{\mbox{$T_{\rm eff}$}}
\newcommand{\logg}{\mbox{$\log g$}}

\newcommand{\vsini}{\mbox{$v\sin i$}}
\newcommand{\kms}{\mbox{km s$^{-1}$}}

\slugcomment{DRAFT}

\begin{document}

\title{\bf MAGNETIC-FIELD MEASUREMENTS OF T TAURI STARS IN THE ORION NEBULA CLUSTER \altaffilmark{1} }

\altaffiltext{1}{Based on observations obtained at the Gemini Observatory, which is operated by the
Association of Universities for Research in Astronomy, Inc., under a cooperative agreement
with the NSF on behalf of the Gemini partnership: the National Science Foundation (United
States), the Science and Technology Facilities Council (United Kingdom), the
National Research Council (Canada), CONICYT (Chile), the Australian Research Council
(Australia), Minist\'erio da Ci\^encia e Tecnologia (Brazil) and SECYT (Argentina). }


\author{Hao Yang \altaffilmark{2}}
\affil{Department of Physics \& Astronomy, Rice University, 6100
Main St. MS-108, Houston, TX 77005} \email{haoyang@colorado.edu}

\author{Christopher M. Johns-Krull} 
\affil{Department of Physics \& Astronomy, Rice University, 6100
Main St. MS-108, Houston, TX 77005} \email{cmj@rice.edu}

\altaffiltext{2}{ Now at JILA, University of Colorado, 440 UCB, Boulder, CO 80309-0440. }

\begin{abstract}

We present an analysis of high-resolution ($R \sim 50,000$) infrared K-band echelle spectra
of 14 T Tauri stars in the Orion Nebula Cluster. We model Zeeman broadening
in three magnetically sensitive \ion{Ti}{1} lines near $2.2\ \mu$m and consistently detect
kilogauss-level magnetic fields in the stellar photospheres. The data are consistent 
in each case with the entire stellar surface 
being covered with magnetic fields, suggesting that magnetic pressure likely dominates
over gas pressure in the photospheres of these stars. These very strong magnetic fields might 
themselves be responsible for the underproduction of X-ray emission of T Tauri stars relative
to what is expected based on main-sequence star calibrations. We combine these results
with previous measurements of 14 stars in Taurus and 5 stars in the TW Hydrae association to
study the potential variation of magnetic-field properties during the first 10 million years
of stellar evolution, finding a steady decline in total magnetic flux with age.

\end{abstract}


\section{INTRODUCTION}

T Tauri stars (TTSs) are a class of low-mass variable stars. Typically a few million years (Myr) old,
TTSs have recently formed out of dense molecular cloud cores and are evolving toward the main sequence along
their Hayashi tracks.
These young solar-type stars display many spectral pecularities, often including excess continuum emission at ultraviolet
through infrared (IR) wavelengths, strong and variable H$\alpha$ and \ion{Ca}{2} H and K emission lines,
both red- and blue-shifted absorption features in Balmer lines, and forbidden line emission. 
These spectral characteristics are most prominent in the spectra of so-called classical TTSs (CTTSs), which 
are TTSs still surrounded by dusty accretion disks. Another category of TTSs, naked TTSs (NTTSs), do not appear 
to have such disks around them and show little or no sign of accretion.
For a comprehensive introduction to TTSs, readers are directed to recent reviews by \citet{menardbertout1999} and
\citet{petrov2003}.

Stellar magnetic fields are thought to play an important role during the evolution of TTSs. Strong magnetic fields
are believed to directly control the interaction between CTTSs and their circumstellar disks \citep[e.g.,][]{bouvier2007}.
Magnetic fields are thought to truncate the disks at or near the corotation radius, forcing disk material
to accrete along the field lines onto the central star \citep{camenzind1990,konigl1991,ccc1993,shu1994,paatz1996}.
While there is abundant evidence in support of this picture \citep{bouvier2007}, observational detection of 
magnetic fields is difficult, and there remain relatively few magnetic field measurements on young stars.

The most successful approach thus far for measuring the total magnetic flux on cool stars has been measuring the 
Zeeman broadening of spectral lines in unpolarized light
\citep{robinson1980, saar1988, valenti1995, cmj1996, cmj1999b,cmj2004,yang2005,yang2008,cmj2007}.
For any given Zeeman component, the splitting resulting from the magnetic field is
$$\Delta\lambda \, [\textrm{m{\AA}}]= {e \over 4\pi m_ec^2} \lambda^2 g B
                = 4.67 \times 10^{-7} \lambda^2 g B \ , \,\,\,\,\,\, 
                     \eqno(1)$$ 
where $g$ is the Land\'e $g$-factor of the transition, $B$ is the strength of the magnetic field in kG, and
$\lambda$ is the wavelength of the transition in \AA. \citet[][hereafter Paper I]{cmj1999b} detected Zeeman broadening
of the \ion{Ti}{1} line at $2.2233\ \mu$m on the CTTS BP Tau and obtained a field strength of $\bar B = 2.6 \pm
0.3$ kG averaged over the entire surface of the star. From the analysis of four \ion{Ti}{1} lines near $2.2\ \mu$m,
\citet[][hereafter Paper II]{cmj2004} measured a strong magnetic field of $2.5 \pm 0.2$ kG on the NTTS Hubble 4.
In addition to studying the \ion{Ti}{1} lines, Paper II also examined several CO lines near $2.3\ \mu$m  which have 
substantially reduced magnetic sensitivity and found no excess broadening beyond that due to stellar rotational and 
instrumental broadening.



Consistent detection of strong fields on TTSs raises the question of the origin of these fields. 
In the largest study to date of magnetic fields on TTSs, \citet{cmj2007} used a sample of 14 stars to look for correlations 
between the measured magnetic-field properties and stellar properties that might be important for dynamo action. 
No significant correlations were found, and \citet{cmj2007} speculated that the fields seen in these young stars may 
be entrained interstellar fields left over from the star formation process \citep[e.g.,][]{tayler1987, moss2003}. 
\citet[][hereafter Papers III and IV]{yang2005,yang2008}
studied six TTSs in the TW Hydrae association (TWA). Roughly 10 Myr old \citep[e.g.,][]{navascues2006}, the TWA stars are 
substantially older than the Taurus stars, whose age is $\sim$ 2 Myr \citep{palla2000}. Paper IV reported
that the total unsigned magnetic flux of the stars appears to decrease with time from 2 Myr to 10 Myr, 
which further supports a primordial origin of the magnetic fields on TTSs.

In this work, we investigate the magnetic properties of TTSs in the Orion Nebula Cluster (ONC), which is even
younger than the Taurus region. At a distance of $400$--$500$ pc \citep{hirota2007,sandstrom2007,jeffries2007}, 
the ONC is one of the most extensively studied regions in the sky. Located just in front of the OMC-1 molecular cloud, 
it is the closest site of massive star formation. The age of the ONC is estimated to be $\sim$ 1 Myr by comparing
the observed HR diagram with theoretical pre-main sequence evolutionary tracks \citep{hillenbrand1997}. 
Strong ionizing radiation, primarily from the young and bright O6 star $\theta^1$ Ori C, evaporates the surrounding 
molecular gas, creating a huge cavity
and exciting the famous Orion Nebula. Though much of this region still suffers strong 
extinction \citep[e.g.,][]{carpenter2001}, 
a few thousand low-mass stars and brown dwarfs are visible in the optical. The young stars in the ONC provide 
an important sample for studies such as of the initial mass function
of stellar clusters, the properties of circumstellar disks, and the evolutionary effects of high-mass
stars on low-mass star formation. 
The dynamic star-formation activity in the ONC gives rise to strong stellar jets and outflows, resulting in
violent interactions with the ambient molecular environment on various scales \citep[e.g.,][]{bally2000}.
\citet{odell2001} gives a comprehensive review of this region.

In this paper, we report magnetic-field measurements for 14 TTSs in the ONC that allow the first
investigation of the magnetic properties of TTSs in this region. In \S\ 2, we describe the observations
and data reduction procedures. In \S\ 3, we present our analysis technique and the magnetic measurements.
A discussion of our results is given in \S\ 4, followed by our conclusions in \S\ 5.

\section{OBSERVATIONS}

We conducted our observations on UT December 10--11, 2006, with the Phoenix high-resolution near-IR
spectrometer \citep{hinkle2003} on the 8.1 m Gemini South Telescope at Cerro Pachon,
Chile (program ID: GS-2006B-C-8). With the K4484 order-sorting filter, the grating was oriented
to provide wavelength coverage between $2.2190$ and $2.2285 \ \mu$m in air.
This configuration allowed the recorded spectra to contain three magnetically sensitive \ion{Ti}{1} lines as 
well as less magnetically sensitive lines of \ion{Fe}{1} and \ion{Sc}{1}.
The 0\farcs34 (4-pixel) wide slit was used to achieve a spectral resolving power of
$R \equiv \lambda/\delta\lambda \sim 50,000$. For each star, two exposures were acquired in nodding
mode with an offset of 5\arcsec\ along the 14\arcsec\ long slit. Typical seeing was around 0\farcs5.
Telluric and spectral standards were also observed each night.
For data reduction, we used custom IDL software developed in Paper I that 
was originally written to handle data from the CSHELL \citep{tokunaga90,greene93} 
spectrometer at the NASA Infrared Telescope Facility.  The
reduction procedures are described in Paper I and required minimal modification
to work with the Phoenix data format.
Seven strong telluric absorption lines were used for wavelength calibration. A summary of the ONC star observations is
provided in Table \ref{oncobs}.


 \begin{table}[!bht]
    \caption {Journal of Observations}\label{oncobs}
\begin{center}
  \leavevmode
  \footnotesize
\resizebox{15.5cm}{!}{
    \begin{tabular}[h]{ccccccc}
    \tableline\tableline
      Object              & R.A. (2000) &  Dec. (2000)   & UT Date     & UT Time  & Total Exposure Time (s) \\[+5pt]
    \tableline
 2MASS 05353126-0518559   & 05:35:31.26 &  -05:24:18.4   & 2006 Dec 10 &  01:56   & 1200                    \\[+5pt]  
 V1227 Ori                & 05:35:10.58 &  -05:21:56.2   & 2006 Dec 10 &  02:40   & 1600                    \\[+5pt]  
 2MASS 05351281-0520436   & 05:35:12.81 &  -05:20:43.6   & 2006 Dec 10 &  03:16   & 1800                    \\[+5pt]  
 V1123 Ori                & 05:35:23.44 &  -05:10:51.7   & 2006 Dec 10 &  03:56   & 2280                    \\[+5pt]  
 OV Ori                   & 05:35:52.76 &  -05:12:59.0   & 2006 Dec 10 &  04:44   & 2600                    \\[+5pt]  
 V1348 Ori                & 05:35:30.42 &  -05:25:38.5   & 2006 Dec 10 &  05:50   & 2880                    \\[+5pt]  
 LO Ori                   & 05:35:08.22 &  -05:37:04.7   & 2006 Dec 10 &  06:49   & 2880                    \\[+5pt]  
 V568 Ori                 & 05:35:36.44 &  -05:34:11.1   & 2006 Dec 10 &  07:45   & 3000                    \\[+5pt]  
 LW Ori                   & 05:35:12.20 &  -05:30:32.9   & 2006 Dec 11 &  01:44   & 3600                    \\[+5pt]  
 V1735 Ori                & 05:35:22.45 &  -05:09:11.0   & 2006 Dec 11 &  02:53   & 3600                    \\[+5pt]  
 V1568 Ori                & 05:35:36.70 &  -05:37:41.5   & 2006 Dec 11 &  04:03   & 3600                    \\[+5pt]  
 2MASS 05361049-0519449   & 05:36:10.49 &  -05:19:44.9   & 2006 Dec 11 &  06:35   & 3600                    \\[+5pt]  
 2MASS 05350475-0526380   & 05:35:04.75 &  -05:26:38.0   & 2006 Dec 11 &  05:15   & 3600                    \\[+5pt]  
 V1124 Ori                & 05:35:47.39 &  -05:13:18.5   & 2006 Dec 11 &  07:46   & 3600                    \\[+5pt]  
    \tableline

  \end{tabular}
}
\end{center}
\end{table}

\section{ANALYSIS AND RESULTS}  \label{section3}

We analyze three \ion{Ti}{1} lines at 2.2211, 2.2233, and 2.2274 $\mu$m using the same procedures that are
described in detail in Papers I--IV. The effective Land\`e g-factors of the three lines are 2.08, 1.66 and
1.58, respectively. First, for our objects, we obtain estimates of the stellar
parameters \Teff, \logg, and \vsini\ from the literature. Then we 
construct model atmospheres appropriate for the stars and calculate synthetic spectra. 
We predict the line profiles with no magnetic broadening and model the observed excess broadening
to measure the surface magnetic fields on the TTSs in Orion.
In contrast to our analysis of the stars in Papers III and IV, we do
not have magnetically insensitive CO data to help verify the accuracy of the atmospheric
parameters. However, two \ion{Fe}{1} lines at 2.2257 and 2.2260 $\mu$m and one
\ion{Sc}{1} line at 2.2267 $\mu$m are available. These three lines have weak sensitivity to magnetic
fields with effective Land\`e g-factors of 1.00, 0.74, and 0.50, respectively. 
So they essentially serve the same purpose as the CO lines (i.e., they help ensure that the excess
broadening seen in the \ion{Ti}{1} lines is magnetic in origin). These three lines are included in the fitting.

In order to construct appropriate model atmospheres, we obtain stellar atmospheric parameters from \citet{hillenbrand1997}.
We convert \citet{hillenbrand1997} spectral types to \Teff\ according to the relation provided in \citet{johnson1966}.
In a few cases, marked in the \Teff\ column in Table \ref{orionpara}, we adjust \Teff\ by 100 or 200 K 
so that the predicted spectrum better matches the \ion{Fe}{1} and \ion{Sc}{1} lines. 
To calculate \logg, we use the stellar luminosities and masses from
\citet{hillenbrand1997} and the \Teff\ converted from spectral types. 
Because of their youth, many of our objects have surface
gravities lower than $3.5$ on the logarithmic scale, which is the lowest available in the NextGen Model
Atmospheres grid \citep{allard95}. In such cases, in order to avoid extrapolating the model atmosphere grid,
we use \logg\ = $3.5$. We expect that doing so 
produces only a modest ($\sim 10\%$) error in our measurements of magnetic fields, because
our analysis technique is not very sensitive to small errors in \Teff\ or \logg, as shown by 
the extensive Mont\'e Carlo analysis in Paper III. 
The \logg\ values we derive from the Hillenbrand paper and those actually used
in the analysis are listed along with other important stellar quantities in Table \ref{orionpara}.

\begin{table}[!bht]
\begin{minipage}{10cm}
  \caption {Stellar Parameters}\label{orionpara}
  \begin{center}
  \leavevmode
  \footnotesize
\resizebox{15.5cm}{!}{
    \begin{tabular}[h]{rccclccrrcccc}
    \tableline\tableline
ID \tablenotemark{a}  &   Object&    & Spectral Type & \Teff & \mbox{$\log g_{H}$}\tablenotemark{b}   & \logg\tablenotemark{c}  & \vsini  & $\log L_{bol}$ & $M_{*}$  & $R_{*}$ & $P_{rot}$\tablenotemark{d} & Age\tablenotemark{e}\\[+5pt]
     &                    &      &               &  (K)   &          &        &  (\kms)   &  ($L_\odot$)  & ($M_\odot$) & ($R_\odot$) & (days) & ($10^6$ yr) \\[+5pt]
    \tableline
 867 & 2MASS 05353126-0518559 & CTTS &  K8    & 3900                 & 2.98 & 3.50 & 11.5$\pm$1.0 &  0.249 & 0.30 & 2.93 &10.66 &  0.7 \\[+5pt]
 347 & V1227 Ori              & CTTS &  K5-K6 & 4200                 & 3.33 & 3.50 & 10.0$\pm$0.8 &  0.086 & 0.41 & 1.72 & 7.33 &  3.4 \\[+5pt]
 391 & 2MASS 05351281-0520436 & CTTS &  M1    & 3500\tablenotemark{f}& 3.27 & 3.50 & 10.6$\pm$1.5 & -0.165 & 0.29 & 2.26 &16.33 &  1.1 \\[+5pt]
 731 & V1123 Ori              & NTTS &  M0/K8 & 3800                 & 3.20 & 3.50 & 16.4$\pm$0.8 &  0.098 & 0.35 & 2.59 & 7.64 &  0.8 \\[+5pt]
1021 & OV Ori                 & CTTS &  K5-K6 & 4000\tablenotemark{f}& 3.62 & 3.62 & 14.0$\pm$1.8 &  0.105 & 0.63 & 2.36 & N/A  &  1.3 \\[+5pt]
 850 & V1348 Ori              & CTTS &  K8    & 3800\tablenotemark{f}& 3.47 & 3.50 & 15.5$\pm$1.3 & -0.107 & 0.40 & 2.04 & 7.78 &  1.3 \\[+5pt]
 321 & LO Ori                 & CTTS &  M0    & 3900\tablenotemark{f}& 3.49 & 3.50 & 12.5$\pm$1.7 & -0.195 & 0.38 & 1.75 & 7.74 &  2.0 \\[+5pt]
 923 & V568 Ori               & CTTS &  M1    & 3664                 & 3.25 & 3.50 &  6.2$\pm$1.9 & -0.192 & 0.29 & 2.00 & 1.15 &  1.3 \\[+5pt]
 381 & LW Ori                 & CTTS &  M0.5  & 3800\tablenotemark{f}& 3.32 & 3.50 & 15.9$\pm$1.2 & -0.150 & 0.32 & 1.95 &16.20 &  1.4 \\[+5pt]
 705 &V1735 Ori\tablenotemark{g}&NTTS&  K4    & 4400\tablenotemark{f}&  N/A & 4.00 & 16.6$\pm$0.8 &   N/A  &  N/A &  N/A & N/A  &  N/A \\[+5pt]
 926 & V1568 Ori              & NTTS &  K7    & 4000                 & 3.29 & 3.50 & 15.0$\pm$1.0 &  0.111 & 0.40 & 2.37 & 5.54 &  1.3 \\[+5pt]
5073 & 2MASS 05361049-0519449 & CTTS &  K3    & 4200\tablenotemark{f}& 3.97 & 4.00 &  9.2$\pm$1.2 &  0.191 & 1.14 & 2.35 & N/A  &  1.4 \\[+5pt]
 258 & 2MASS 05350475-0526380 & CTTS &  M0.5  & 3664                 & 3.34 & 3.50 & 10.4$\pm$1.0 & -0.157 & 0.33 & 2.08 &10.98 &  1.2 \\[+5pt]
 995 & V1124 Ori              & NTTS &  M1.5  & 3589                 & 3.17 & 3.50 &  6.2$\pm$1.3 & -0.162 & 0.26 & 2.15 & 8.18 &  1.2 \\[+5pt]
    \tableline
    \tableline

\tablenotetext{a}{The ID, spectral type, $L_{bol}$, and $M_{*}$ are taken from \citet{hillenbrand1997}. The \vsini\ is from \citet{sicilia2005}.}
\tablenotetext{b}{Calculated using $L_{bol}$, $M_{*}$ from \citet{hillenbrand1997} and \Teff\ used in our models. See text in \S\ 3.}
\tablenotetext{c}{Used for this analysis.}
\tablenotetext{d}{Rotation periods are taken from \citet{flaccomio2003a}.}
\tablenotetext{e}{Stellar ages are calculated from the pre-MS evolutionary tracks of \citet{siess2000}.}
\tablenotetext{f}{\Teff\ is adjusted by 100 or 200 K after converted from spectral type. See text in \S\ 3.}
\tablenotetext{g}{Bolometric luminosity of V1735 Ori is not available in the literature, 
                  so there are no estimates of radius, surface gravity, mass and age for this star.}

  \end{tabular}
}
 \end{center}
 \end{minipage}
\end{table}

The magnetic models we use for the ONC stars are constructed in the same way as models 1--3 for the TWA stars in
Paper IV, where they are fully described. Briefly, the stellar surface is treated as being composed of several regions. 
Each region has a filling factor (i.e., a fractional size) and is covered with a single magnetic field strength. 
The sum of all filling factors must be unity. All the models used here include a field-free region ($B = 0$); however,
Papers I and II show that generally the data do not demand the models have such a field-free region on the stellar surface.
We also assume that these regions are spatially well mixed over the surface of the star. Models 1, 2, and 3 allow one,
two, and three magnetic regions on the stellar surface, respectively, in addition to the field-free region. 
For models 1 and 2, the field strengths are free parameters.  For model 3, we have field strengths in the three magnetic 
regions fixed at 2 kG, 4 kG, and 6 kG and fit only for the filling factor of each region. 
The mean magnetic field is just the sum of the field strengths ($B_i$) in the 
magnetic regions weighted by their corresponding filling factors ($f_i$), i.e., $\bar B = \displaystyle\sum_{i=1}^n B_if_i$,
where $n$ is the number of magnetic regions on the stellar surface in each model.

To compute the model spectrum for a star,
we first interpolate the NextGen Model Atmosphere grid to the appropriate \Teff\ and \logg\ and construct a model
atmosphere specific for each star. Then we compute the spectrum for each magnetic-field region using the model atmosphere 
and a polarized radiative transfer code \citep{piskunov1999} that assumes a radial     
magnetic geometry in the stellar photosphere. We add in rotational broadening to the line profiles according to the \vsini\
of the star and convolve the resulting profiles with a Gaussian function corresponding to the observed spectral resolution ($50,000$). 
The spectrum for each region is then weighted by its filling factor and summed to form the final model spectrum. 
We use the nonlinear least-squares technique of Marquardt \citep{bevington1992} and solve for the combination of field 
strengths and filling factors that yields the best fit to the observed spectrum.

The majority of our objects are CTTSs and are subject to substantial K-band veiling. Veiling is an excess continuum source, 
presumably from the circumstellar disk, that acts to weaken lines in continuum normalized spectra.
Therefore, in addition to solving for magnetic field strengths
and filling factors, our models also allow K-band veiling ($r_{K}$) as a free parameter and assume it to be constant
across the wavelength range of interest. 
This extra parameter only affects the depths of the lines and does not directly change the
line widths. The \ion{Fe}{1} and \ion{Sc}{1} lines help ensure that the veiled
\ion{Ti}{1} line profiles with little magnetic broadening are accurately predicted. The measured
mean magnetic field for each of the magnetic models (models 1--3), fitted K-band veiling, and corresponding
reduced $\chi^2$ values are listed in Table \ref{orionresults}. Typical uncertainty in the veiling measurements is 
between 0.05 and 0.10.

\begin{figure}[ht]
  \begin{center}
    \includegraphics[scale=0.45]{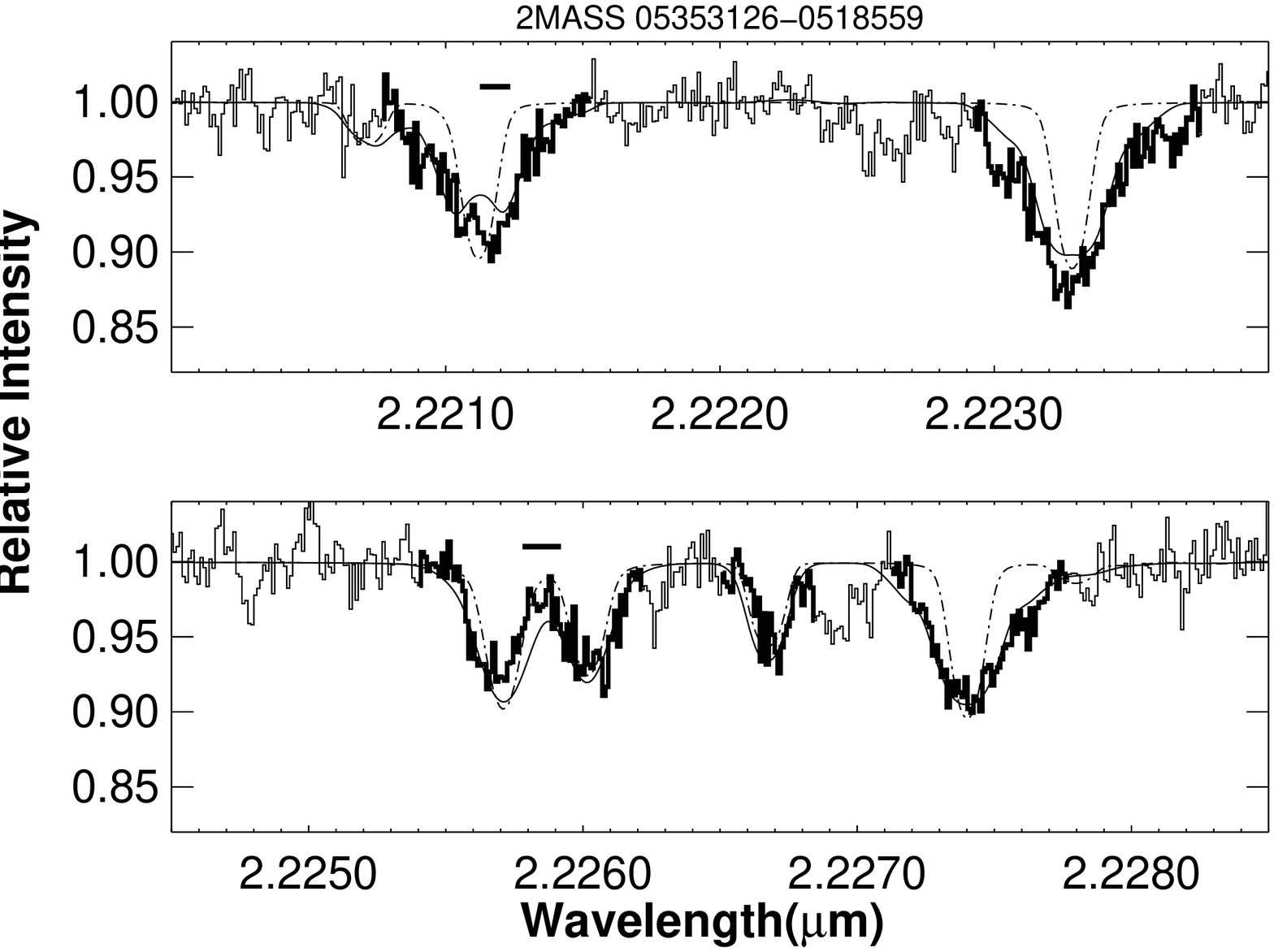}
   \caption { K-band infrared spectra of 2MASS 05353126-0518559 are shown in the histogram.
        The spectral regions shown in bold are the magnetically sensitive \ion{Ti}{1} lines actually
        used in the fit. The dash-dotted lines show a model with no magnetic field.
        The smooth lines are the best fit with magnetic broadening based on model 3.
        The solid horizontal bars mark the wavelengths where telluric absorption is stronger than 3\% of the continuum and
        coincides with the lines used in the fit.}
           \label{orionplot02}
   \end{center}
\end{figure}

\begin{figure}[ht]
  \begin{center}
    \includegraphics[scale=0.45]{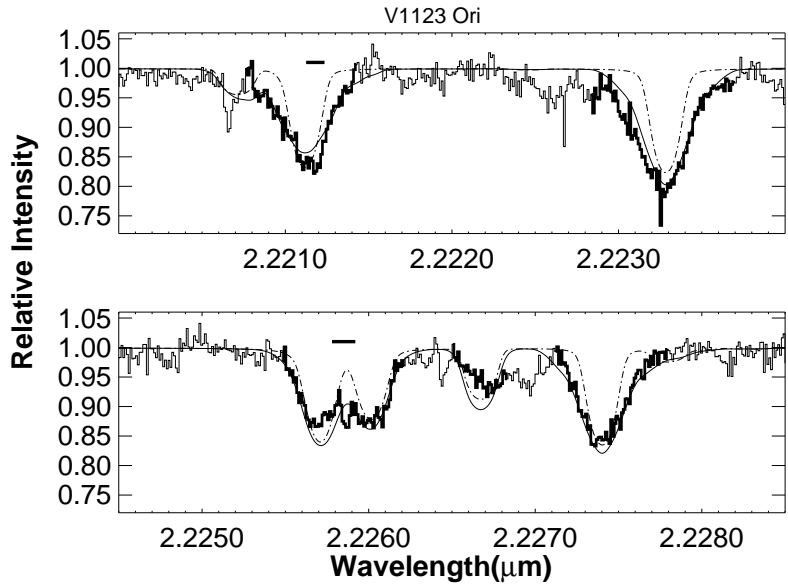}
   \caption { K-band infrared spectra of V1123 Ori are shown in the histogram.
        The spectral regions shown in bold are the magnetically sensitive \ion{Ti}{1} lines actually
        used in the fit.
        The magnetically sensitive Ti I lines are in bold.
        The dash-dotted lines show a model with no magnetic field.
        The smooth lines are the best fit with magnetic broadening based on model 3. 
        The solid horizontal bars mark the wavelengths where telluric absorption is stronger than 3\% of the continuum and
        coincides with the lines used in the fit.}
           \label{orionplot05}
   \end{center}
\end{figure}

\begin{figure}[ht]
  \begin{center}
    \includegraphics[scale=0.45]{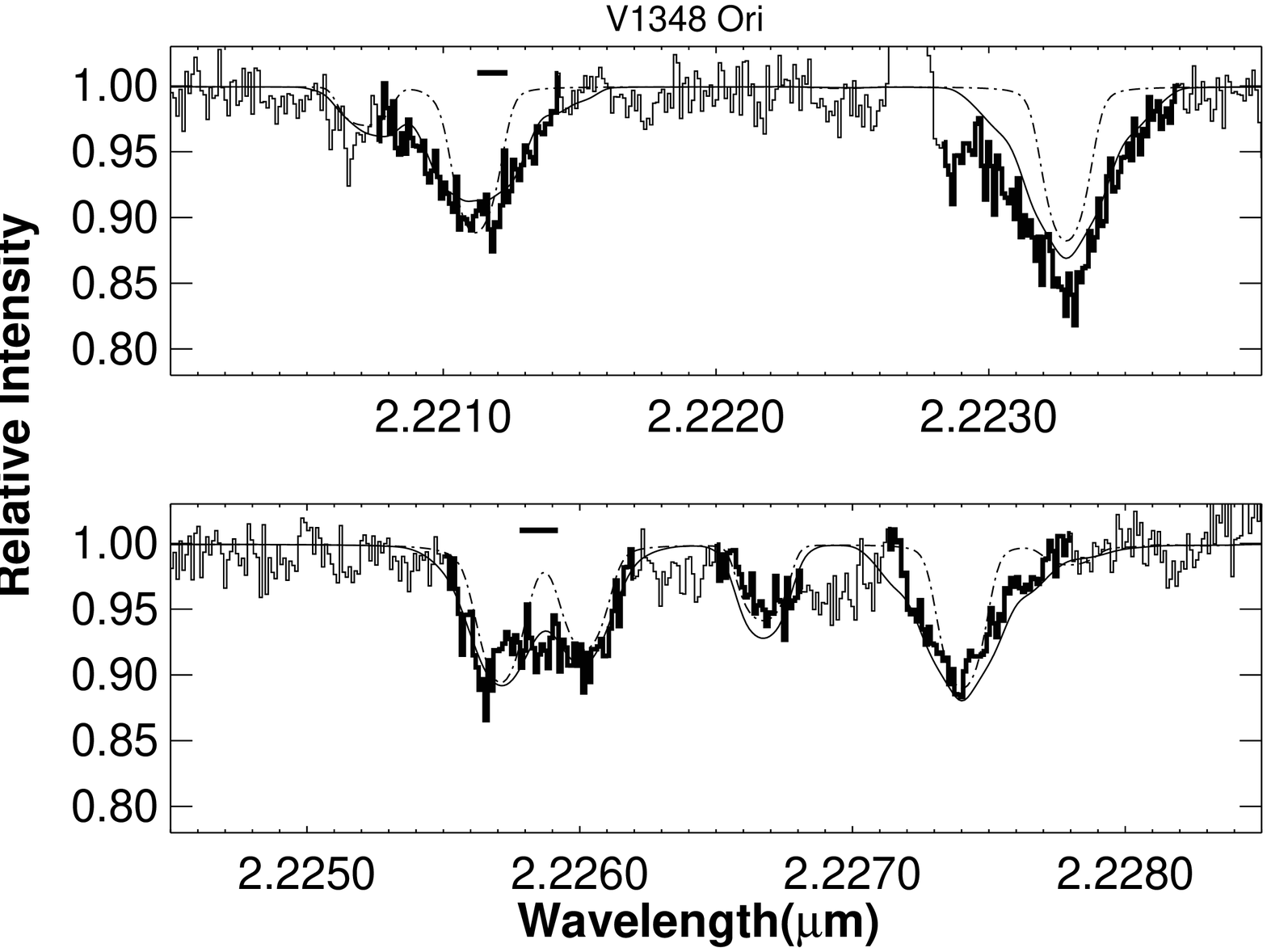}
   \caption { K-band infrared spectra of V1348 Ori are shown in the histogram.
        The spectral regions shown in bold are the magnetically sensitive \ion{Ti}{1} lines actually
        used in the fit. The dash-dotted lines show a model with no magnetic field.
        The smooth lines are the best fit with magnetic broadening based on model 3. 
        The solid horizontal bars mark the wavelengths where telluric absorption is stronger than 3\% of the continuum and
        coincides with the lines used in the fit. The emission feature at 2.2227 $\mu$m is a molecular hydrogen line.}
           \label{orionplot08}
   \end{center}
\end{figure}

As examples of the fits, the K-band spectra of 2MASS 05353126-0518559,
V1123 Ori, and V1348 Ori are shown in Figures~\ref{orionplot02}--\ref{orionplot08}. (The spectra of other targets 
are presented as \emph{online-only} Figures 7-17.)
When correcting the observed spectra for telluric absorption, there are only two particularly strong telluric lines that affect 
the photospheric lines used in our analysis.
The horizontal bars in Figures~\ref{orionplot02}--\ref{orionplot08} mark the two telluric features that are stronger than 3\% of the continuum.
Though these two regions are likely to have higher uncertainties than the regions that not affected by telluric absortion,
\citet{cmj2009} found that arbitrarily increasing the uncertainties of the
pixels in Phoenix data that are significantly affected by the telluric 
correction has essentially no effect on the fitting results.  The Phoenix
data used here is even less affected by telluric contamination and we
expect uncertainties in its correction to have similarly negligible effect.
To show an example of how sensitive our analysis is to the uncertainties in \Teff, we analyze 2MASS 05353126-0518559 
with a \Teff\ that is $\pm\ 300$ K from the value we used (3900 K), and 
the measured magnetic field values are changed by only 3\% ($- 300$ K) and 8\% ($+ 300$ K).
As demonstrated in Paper III, small errors in \Teff\ and \logg\ in general only affect the magnetic field measurements by a few percent.

Additional errors might be incurred by errors in \vsini\ values, as we do not
have magnetically insensitive CO lines to check on the atmospheric 
parameters (in particular \vsini ) used in the spectrum synthesis.
However, the lack of CO line data is not expected to be a major issue 
for this work.  In previous papers in this series, the CO data
have largely served as a check on optically determined \vsini\
measurements. For example, \citet{cmj2007} did use \vsini\ measurements
determined from the CO data for most stars in that study.  These \vsini\ 
values were published in \citet{cmj2001} where it was shown that the CO
determined \vsini\ values are consistent with optically 
determined values such as those used in the present study.
To estimate the effect of uncertainties in \vsini\ on the recovered
values of $\bar B$, we have repeated
the analysis of 2MASS 05353126-0518559 (medium \vsini, strong
field) and V568 Ori (low \vsini, weaker field) where we have held
the vsini of each star fixed at values corresponding to $\pm 1$ km s$^{-1}$
(a typical uncertainty).  In the case of the 2MASS star, the
resulting change in $\bar B$ was 1\% and 3\%, and in the case of V568 Ori
the resulting change was 1.5\% and 8\%.  
The reason for such a small effect of the recovered field 
is that the magnetic broadening is significantly stronger
than the rotational broadening in the \ion{Ti}{1} lines.  To make a
final estimate of our field uncertainties from possible inaccuracies in stellar parameters, 
we conservatively assume
a 10\% uncertainty resulting from mismatches in both \Teff\ and \logg\,
and add these together in quadrature with an 8\% uncertainty resulting
from errors in the adopted \vsini. These sources of uncertainties add up to 16 \%.
For most of our stars, there is a noticeable decrease in $\chi^2$ for models
2 and 3 relative to the single-component model (1).  The two stars for which
this is not the case are OV Ori and LW Ori, which both have relatively low
signal-to-noise spectra.  On the other hand, models 2 and 3 do not agree
perfectly, even though they do generally return mean fields that agree to
within the 16\% uncertainty described above.  Therefore, we adopt as our
final mean field value the average mean field returned by models 2 and 3.
We take the difference between this mean and the individual values as an
estimate of the uncertainty of the technique, and add it in quadrature to
the 16\% uncertainty described above.  For OV Ori and LW Ori, we average
the mean field from all 3 models and take the standard deviation as the
an estimate of the uncertainty, which is added in quadrature to the 16\%
uncertainty.
The final $\bar B$ values and the associated uncertainties are listed 
in the last two columns of Table \ref{orionresults}.
We use the adopted values of $\bar B$ and its uncertainty, and 
propagate them through all subsequent calculations and comparisons.

\begin{table}[!bht]
  \caption{Magnetic-Field Measurements.} \label{orionresults}
  \begin{center}
\resizebox{15cm}{!}{
   \begin{tabular}[h]{ccrccccccccccc}
   \tableline\tableline
Object&Type& &\multicolumn{3}{c}{Model 1~~~~}   & \multicolumn{3}{c}{Model 2~~~~}  & \multicolumn{3}{c}{Model 3~~~}  & \multicolumn{2}{c}{Adopted}              \\ \cline{4-5}  \cline{7-8} \cline{10-11} \cline{13-14}
      &   & $\chi_r^2$ (B=0) &  $\bar B $ (kG) ($\chi_r^2$) & $r_{K}$  &       & $\bar B $     (kG) ($\chi_r^2$) &  $r_{K}$  &  & $\bar B$ (kG) ($\chi_r^2$) &  $r_{K}$ & & $\bar B$ (kG) & $\sigma$ (kG)   \\
\tableline
2MASS 05353126-0518559 & CTTS &14.19& 1.91 (1.99) & 0.85 & & 2.68 (1.60)  & 1.03& & 2.99 (1.62)  &  1.05 && 2.84 & 0.48\\[+5pt]
V1227 Ori              & CTTS &11.99& 1.81 (3.00) & 0.72 & & 2.13 (2.50)  & 0.62& & 2.15 (2.77)  &  0.65 && 2.14 & 0.34\\[+5pt]
2MASS 05351281-0520436 & CTTS &17.85& 1.41 (1.85) & 0.63 & & 1.81 (1.81)  & 0.81& & 1.58 (1.82)  &  0.66 && 1.70 & 0.29\\[+5pt]
V1123 Ori              & NTTS & 9.71& 1.94 (3.09) & 0.00 & & 2.43 (2.64)  & 0.00& & 2.58 (2.69)  &  0.00 && 2.51 & 0.41\\[+5pt]
OV Ori                 & CTTS & 3.16& 1.55 (1.10) & 0.59 & & 1.64 (1.19)  & 0.63& & 2.35 (1.06)  &  0.72 && 1.85 & 0.53\\[+5pt]
V1348 Ori              & CTTS & 9.78& 2.05 (3.54) & 0.38 & & 3.05 (3.14)  & 0.52& & 3.23 (2.74)  &  0.58 && 3.14 & 0.51\\[+5pt]
LO Ori                 & CTTS & 9.45& 2.70 (2.81) & 1.44 & & 3.26 (2.52)  & 1.61& & 3.63 (2.41)  &  1.56 && 3.45 & 0.58\\[+5pt]
V568 Ori               & CTTS &10.98& 1.32 (1.83) & 0.64 & & 1.59 (1.30)  & 0.65& & 1.46 (1.39)  &  0.55 && 1.53 & 0.25\\[+5pt]
LW Ori                 & CTTS & 8.30& 1.04 (1.32) & 1.43 & & 1.32 (1.33)  & 1.56& & 1.55 (1.30)  &  1.58 && 1.30 & 0.33\\[+5pt]
V1735 Ori              & NTTS & 3.12& 1.54 (1.47) & 0.00 & & 2.14 (1.38)  & 0.00& & 2.01 (1.38)  &  0.00 && 2.08 & 0.34\\[+5pt]
V1568 Ori              & NTTS & 4.00& 1.06 (2.16) & 0.00 & & 1.42 (1.77)  & 0.00& & 1.42 (2.01)  &  0.00 && 1.42 & 0.23\\[+5pt]
2MASS 05361049-0519449 & CTTS & 9.99& 1.66 (2.10) & 0.24 & & 2.28 (1.71)  & 0.26& & 2.34 (1.67)  &  0.25 && 2.31 & 0.37\\[+5pt]
2MASS 05350475-0526380 & CTTS &13.95& 1.97 (2.90) & 0.60 & & 2.81 (2.42)  & 0.79& & 2.76 (2.44)  &  0.70 && 2.79 & 0.45\\[+5pt]
V1124 Ori              & NTTS & 8.90& 1.45 (3.90) & 0.00 & & 2.17 (3.19)  & 0.00& & 2.00 (3.25)  &  0.00 && 2.09 & 0.34\\[+5pt]
 
\tableline
\end{tabular}
}
\end{center}
\end{table}

\section{DISCUSSION}

For 14 T Tauri stars in the ONC, we model the Zeeman broadening in three 
magnetically sensitive \ion{Ti}{1} lines and measure the mean surface magnetic
field strengths on these stars.  We have attempted to account for various
sources of uncertainty in our measurements, but there are some potential
sources for which this is quite difficult.  An example of this is the
assumed geometry of the field.  We have assumed a purely radial field
geometry in the spectrum synthesis.  This was originally motivated by
analogy to the Sun where observations suggested surface fields (at least
in high flux density regions) are primarily oriented along the vertical direction
\citep[e.g.,][]{stenflo2010}.  A number of recent Zeeman Doppler
Imaging (ZDI) studies have begun to provide surface maps of magnetic fields
on T Tauri stars \citep[e.g.,][]{donati2007,donati2008a,donati2010,hussain2009}. 
These studies show that the field topology is not purely radial;
however, it is not at all clear that the ZDI results give an accurate
picture of the majority of the magnetic flux on the star.  \citet{reiners2009} 
showed that in strongly magnetic M stars, Stokes V based
studies (ZDI) miss 86\% or more of the magnetic flux present compared
to Stokes I based studies due to the effects of flux cancellation in
Stokes V measurements which are not present in Stokes I measurements. 

Our work here is based on Stokes I.  If the field were entirely
aimed along the line of sight, or entirely perpendicular to it,
the relative strength of the $\pi$ and $\sigma$ Zeeman components would 
change and affect our results somewhat.  However, the  polarization
measurements mentioned above show such pathological field
orientations are generally not present \citep[see also][]{cmj1999a, daou2006,yang2007}.
Instead, there is likely a 
stong mixture of field orientations on the star: some parallel to the line
of sight, some perpendicular to the line of sight, and all angles
in between.  While we assume a radial geometry on the star, when
we integrate over the visible stellar disk, we then include a
good mixture of field lines with all orientations to the line of sight.
As a result, we do not expect our results to be strongly affected by our
assumption of a radial field geometry.  Again, as mentioned above, since
the ZDI results likely do not provide good guidance for the geometry of 
the total field \citep{reiners2009}, this is probably the best we can
do until better guidance comes from theory or even better observations.

For most of our targets, the single magnetic component fit of model 1 is not good enough to account for all the
excess broadening in the observed line profiles. For a few targets such as OV Ori and LW Ori, adding more magnetic 
components does not improve the fits significantly. That is due to a combination of low signal-to-noise ratios 
and small magnetic broadening.
Models 2 and 3 generally yield higher mean magnetic field strengths and better matches to the observations, resulting in lower reduced $\chi^2$ values. 
In most cases, the mean field values recovered by models 2 and 3 are in good agreement with
similar reduced $\chi^2$ values. On average, the difference between the two recovered values is less than 9\%.

\begin{figure}[!bht]
  \begin{center}
    \includegraphics[scale=0.60]{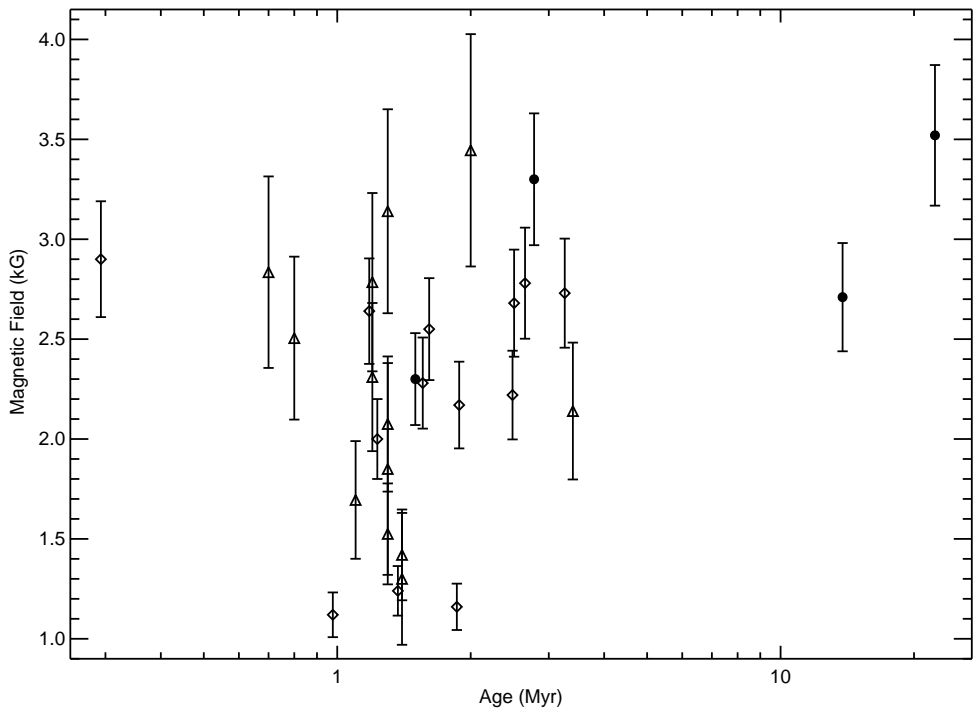}
   \caption[Measured Magnetic Fields Plotted Against Age
         For the Orion, Taurus/Auriga and TWA Stars ]
        { Measured magnetic fields plotted against age for the Orion, Taurus/Auriga and TWA Stars.
          The triangles represent the Orion stars from this work. The filled circles represent
         the TWA stars from Papers III and IV, and the hollow diamonds are
         the Taurus/Auriga stars from \citet{cmj2007}. The sizes of error bars are 10\% for the TWA and Taurus/Auriga stars
         and the values in the last column of Table \ref{orionresults} for the Orion stars. }
           \label{bfieldage}
   \end{center}
\end{figure}

\begin{figure}[!bht]
  \begin{center}
    \includegraphics[scale=0.60]{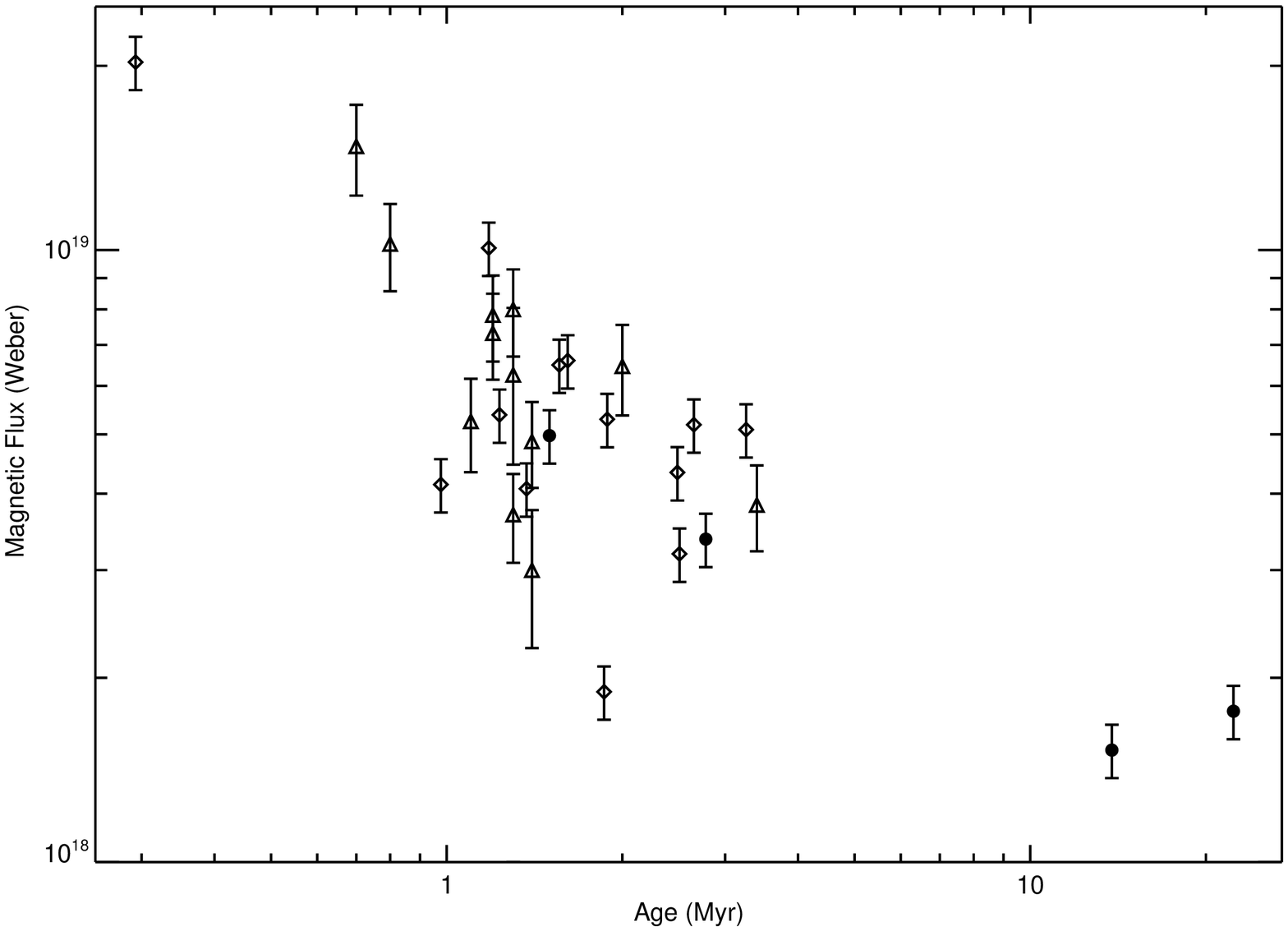}
   \caption[Measured Magnetic Flux Plotted Against Age
         For the Orion, Taurus/Auriga and TWA Stars ]
        { Measured magnetic flux plotted against age for the Orion, Taurus/Auriga and TWA Stars.
          The triangles represent the Orion stars from this work. The filled circles represent
         the TWA stars from Papers III and IV, and the hollow diamonds are
         the Taurus/Auriga stars from \citet{cmj2007}. }
           \label{fluxage}
   \end{center}
\end{figure}

Since we apply models that are consistent with previous studies of 14 stars in Taurus \citep{cmj2007} and 
5 stars in the TW Hydrae association (Papers III and IV), we are able to directly compare the measurements of 
all 33 stars in the three regions. Out of the 33 stars in total, 27 stars have spectral types between K7 and M3. 
The rest of the sample have spectral types between K3 and K6, except for the K0 star T Tau. Therefore, these stars
possess similar spectral properties and allow a meaningful comparison.
We calculate stellar ages using the pre-MS evolutionary tracks of \citet{siess2000}, and plot the measured 
magnetic field values and magnetic fluxes ($4\pi {R_*}^2\bar B$) against stellar ages, shown in Figures \ref{bfieldage} 
and \ref{fluxage}, respectively.
We note that while stellar magnetic field strength spans a range of values and displays no apparent correlation with
stellar age, stellar magnetic flux decreases consistently with time.
The primary reason for the inverse correlation of magnetic flux with
  age is the diminishing stellar radii as the stars evolve down their
  Hyashi tracks.  However, as shown below, the measured fields/fluxes
  do not correlate with any other expected dynamo properties, including
  for example the stellar gravity.  As a result, we find no evidence
  the the field strengths are being set by some other property of the
  stars, which might produce an inverse correlation with age/radius as
  a by-product. 
The observed trend of the flux decreasing with age supports the idea that the strong magnetic fields of TTSs could be of
primordial origin. As suggested by \citet{moss2003}, during the early stages of stellar evolution, while
primordial magnetic flux does decay, young stars may be able to maintain part of the fossil flux
from the molecular cloud material, especially later when a raditive core is formed. It should be noticed that
currently there are only a few measurements for stars younger than 1 Myr or older than 10 Myr, so more observations
are needed to further trace out the evolution of stellar magnetic fields with time.

In addition to a fossil origin for the magnetic fields on young stars, these observed strong fields
could in principle be the result of dynamo action, either a solar-like $\alpha-\Omega$ dynamo \citep{durney1978,parker1993}
or as a result of a distributed, turbulent or convective dynamo \citep[e.g.,][]{durney1993}.
In current $\alpha-\Omega$ dynamo models, magnetic fields are generated by differential rotation at the region between
the radiative interior and the outer convective shell, a region known as the tachocline. 
In such a dynamo, it is believed that levels of magnetic activity are
related to the Rossby number, defined as $P_{rot}/\tau_{c}$, where $P_{rot}$ is the rotation period
and $\tau_{c}$ is the convective turnover time. However, \citet{cmj2007} found no correlation for 14
Taurus stars and TW Hya between the measured magnetic field values and various stellar or dynamo parameters. 
Here, for a combined sample of 31 stars from \citet{yang2008}, \citet{cmj2007} and this study, we look for a correlation
between measured magnetic properties and various potential dynamo parameters. We obtain $\tau_{c}$ from models described 
in \citet{kim1996} and \citet{preibisch2005}. In Table~\ref{correl}, we give the Pearson linear correlation coefficient, 
$r$, and its false alarm probability, $f_{p}$, for several comparisons. No significant correlation is seen between the observed
magnetic field strength/flux and any dynamo parameter. 
Such lack of correlation is generally contrary to the prediction of
theories of solar-like interface dynamos. Besides interface dynamos, there are models that involve a distributed 
dynamo \citep[e.g.,][]{durney1993}.
This kind of dynamo is thought to be able to operate in fully convective stars such as TTSs and create
large-scale magnetic fields via convective motions. Recently, \citet{christensen2009} extended a scaling law derived
from convection-driven geodynamo models to the regime of stars, including TTSs, and
their predictions for the field strengths on TTSs are generally consistent with the observed values.
However, magnetic fields generated by a distributed dynamo are also
expected to have a dependence, though possibly weak, on rotation\citep{durney1993,chabrier2006}, which we do not observe. 
Therefore, we suggest that the survival of the fossil magnetic flux seems to be a more promising explanation
for the origin of strong magnetic fields on TTSs; however, more work in this area is needed.

\begin{table}[!bht]
\caption{Correlations of Measured Magnetic Parameters with Dynamo Parameters}
\label{correl}
\begin{center}
\begin{tabular}{lrr}
 \tableline\tableline
Quantities Compared                      &  $r$    & $f_{p}$ \\[+5pt]
\tableline
$\bar B$ vs. $P_{rot}^{-1}$              &  -0.01  &  0.94   \\[+5pt]
$\Phi$ vs. $P_{rot}^{-1}$                &  -0.04  &  0.83   \\[+5pt]
$\bar B$ vs. $\tau_{c}$                  &   0.24  &  0.20   \\[+5pt]
$\Phi$ vs. $\tau_{c}$                    &  -0.08  &  0.64   \\[+5pt]
$\bar B$ vs. $\tau_{c}P_{rot}^{-1}$      &   0.05  &  0.79   \\[+5pt]
$\Phi$ vs. $\tau_{c}P_{rot}^{-1}$        &  -0.15  &  0.42   \\[+5pt]
$\bar B$ vs. \logg                       &   0.14  &  0.46   \\[+5pt]
$\Phi$ vs. \logg                         &  -0.45  &  0.01   \\[+5pt]
\tableline
\end{tabular}
\end{center}
\end{table}

\begin{figure}[ht]
  \begin{center}
    \includegraphics[scale=0.60]{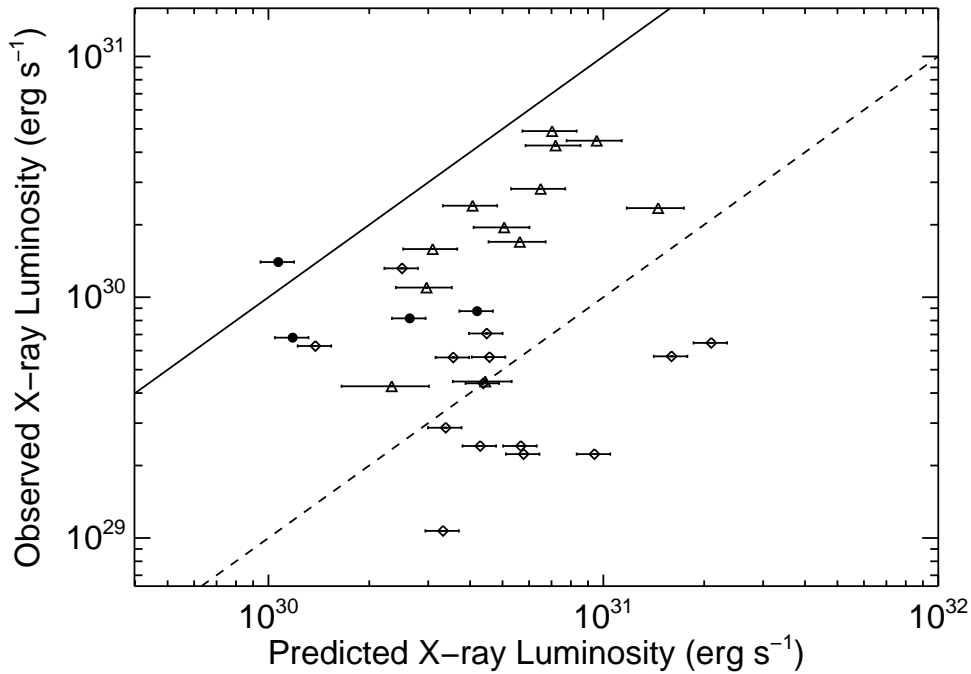}

   \caption{Observed X-ray luminosity plotted against the predicted
         X-ray luminosity for the Orion, Taurus/Auriga and TWA stars.
         The solid line is the line of equality, and the dashed line is
         where the predicted value is 10 times the observed one. The triangles
         represent the Orion stars from this work. The filled circles represent
         the TWA stars from Paper IV, and the hollow diamonds are
         the Taurus/Auriga stars from \citet{cmj2007}. }

           \label{orionxrayplot}
   \end{center}
\end{figure}

We also examine the relationship between magnetic flux and the X-ray properties of the ONC stars.
As first noted by \citet{pevtsov2003} for the Sun and a number of main-sequence dwarfs, there is a
nearly linear correlation between X-ray luminosities, $L_{\rm x}$, and the total unsigned magnetic
flux, $\Phi$. The correlation holds over 12 orders of magnitude in both quantities. We use this relationship with our measured
magnetic flux values to calculate the predicted X-ray luminosities for our ONC stars. These values are listed
in Table \ref{orionxray}, where we compare them with the observed values reported by
\citet{flaccomio2003a} and \citet{getman2005a}. The typical uncertainty in the observed $L_x$ is 0.2 dex. 
The comparison is also plotted in Figure \ref{orionxrayplot}.
On average, the predicted $L_x$ values of the ONC stars are larger than the observed ones by a
factor of $4.9 \pm 1.4$. This suppression of the X-ray emission is consistently found as well for the
Taurus and TWA stars, also shown in Figure \ref{orionxrayplot}, though the reduction is different in each region.

Large flares are thought to be responsible for harder X-ray emission on TTSs, while softer, persistent X-ray emission may
be generated by so-called microflares that
release energy from stressed magnetic fields in the coronae of these stars \citep{guedel2003,arzner2007}.
On the Sun, these stresses are believed to be built up by convective motions acting on the footpoints of coronal loops. 
For TTSs, strong surface magnetic fields may decrease the efficiency by which convective gas motions are able to move 
footpoints around and build up these magnetic
stresses. Thus, in addition to listing the measured magnetic field strengths based on model 3, $\bar B$, 
we also list in Table \ref{orionxray} the equipartition field strength, $B_{eq}={(8\pi P_{g})}^{1/2}$. 
This quantity is calculated assuming that the magnetic pressure, ${B_{eq}}^2/(8\pi)$, is
in balance with the surrounding unmagnetized photospheric gas pressure, $P_{g}$. We use the gas pressure from 
a location in the model atmospheres where the local gas temperature is approximately equal to the effective temperature
of the star. This level in the stellar photosphere is in the vinicity of where the observed continuum forms, so the gas pressure
here should be generally greater than the regions where the \ion{Ti}{1} lines form, setting an upper limit.
The ratio of the
observed field strength to the equipartition field strength ranges from 1.3 to 3.9 for the ONC
stars, indicating the dominance of magnetic pressure over gas pressure in the photospheres of these TTSs.
This dominance of magnetic pressure may partially prohibit the production of X-ray emission. 
Since these ratios of the ONC stars are close to those of the TWA stars, it is likely no coincidence that in 
Figure \ref{orionxrayplot} the observed X-ray luminosities of most ONC stars are smaller than the predicted
values by a similar factor as those of the TWA stars.

\begin{table}[!bht]
  \caption{Magnetic-Field Properties and X-Ray Luminosities.} \label{orionxray}
  \begin{center}
\resizebox{14cm}{!}{
   \begin{tabular}[h]{ccccc}
   \tableline \tableline
  Object               & $ \bar B$      & $ B_{eq}$  & Predicted $\log L_X$  & Observed $\log L_X$  \\[+5pt]
                       &   (kG)         &   (kG)     &     (erg/s)           & (erg/s)              \\[+5pt]
\tableline
2MASS 05353126-0518559 &    2.84            &  0.99              & 31.16   &  30.37   \\[+5pt]
V1227 Ori              &    2.14            &  0.87              & 30.49   &  30.20   \\[+5pt]
2MASS 05351281-0520436 &    1.70            &  1.03              & 30.65   &  29.65   \\[+5pt]
V1123 Ori              &    2.51            &  1.00              & 30.98   &  30.65   \\[+5pt]
OV Ori                 &    1.85            &  1.07              & 30.73   &  N/A     \\[+5pt]
V1348 Ori              &    3.14            &  1.00              & 30.86   &  30.63   \\[+5pt]
LO Ori                 &    3.45            &  0.99              & 30.75   & $<$30.23   \\[+5pt]
V568 Ori               &    1.53            &  1.07              & 30.47   &  30.04   \\[+5pt]
LW Ori                 &    1.30            &  1.00              & 30.37   &  29.63   \\[+5pt]
V1735 Ori              &    2.08            &  1.35              &  N/A    &   N/A    \\[+5pt]
V1568 Ori              &    1.42            &  0.93              & 30.61   & $<$30.38   \\[+5pt]
2MASS 05361049-0519449 &    2.31            &  1.42              & 30.85   &  30.69   \\[+5pt]
2MASS 05350475-0526380 &    2.79            &  1.02              & 30.81   &  30.45   \\[+5pt]
V1124 Ori              &    2.09            &  1.67              & 30.70   &  30.29   \\[+5pt]

\tableline
\end{tabular}
}
\end{center}
\end{table}

\section{Conclusion}

By modeling the Zeeman broadening in three magnetically sensitive \ion{Ti}{1} lines,
we consistently measure kilogauss-level magnetic fields in the photospheres of 14 TTSs 
in the ONC. We find a systematic decrease of stellar magnetic flux from the young Orion
region ($\sim$ 1 Myr), through the Taurus region ($\sim$ 2 Myr), to the TWA ($\sim$ 10 Mys).
This finding supports a primordial origin of the magnetic fields on TTSs, though recent convective
dynamos by \citet{christensen2009} are able to produce magnetic fields that have strengths consistent 
with observations.
Because convective
motions are likely partially inhibited by the dominance of magnetic pressure over unmagnetized 
gas pressure in the stellar photospheres, the strong magnetic fields could also be responsible for 
suppression of X-ray emission on TTSs relative to that expected from main sequence star 
calibrations.

\acknowledgments
This paper is based on observations obtained with the Phoenix infrared
spectrograph, developed and operated by the National Optical Astronomy
Observatory.

\appendix
\section{Online-Only Figures}

\begin{figure}[ht]
  \begin{center}
    \includegraphics[scale=0.45]{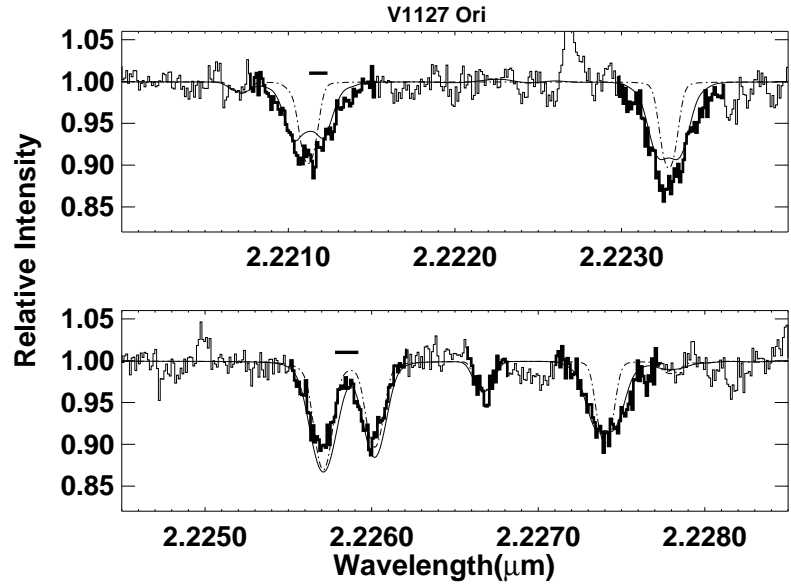}
   \caption {K-band infrared spectra of V1227 Ori are shown in the histogram.
        The spectral regions shown in bold are the magnetically sensitive \ion{Ti}{1} lines actually
        used in the fit. The dash-dotted lines show a model with no magnetic field.
        The smooth lines are the best fit with magnetic broadening based on model 3.
        The solid horizontal bars mark the wavelengths where telluric absorption is stronger than 3\% of the continuum and
        coincides with the lines used in the fit.}
           \label{plot03}
   \end{center}
\end{figure}

\begin{figure}[ht]
  \begin{center}
    \includegraphics[scale=0.45]{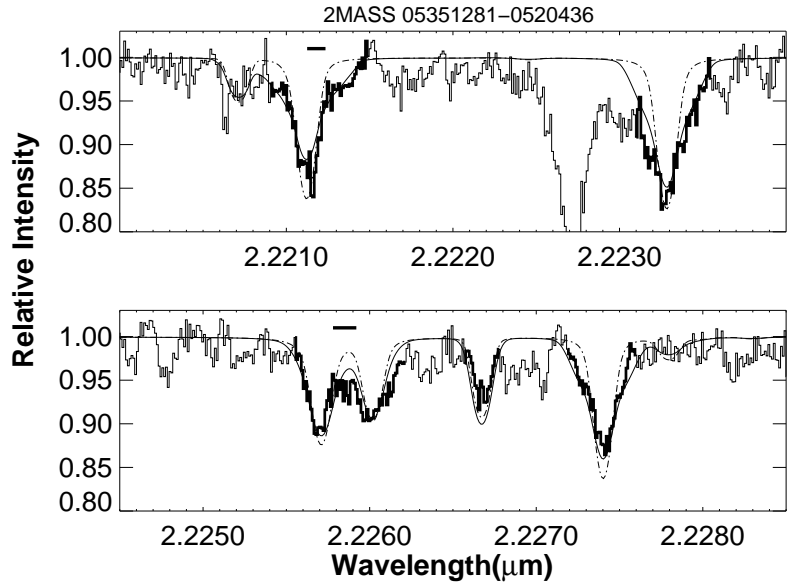}
   \caption {K-band infrared spectra of 2MASS 05351281-0520436. }
           \label{plot04}
   \end{center}
\end{figure}

\begin{figure}[ht]
  \begin{center}
    \includegraphics[scale=0.45]{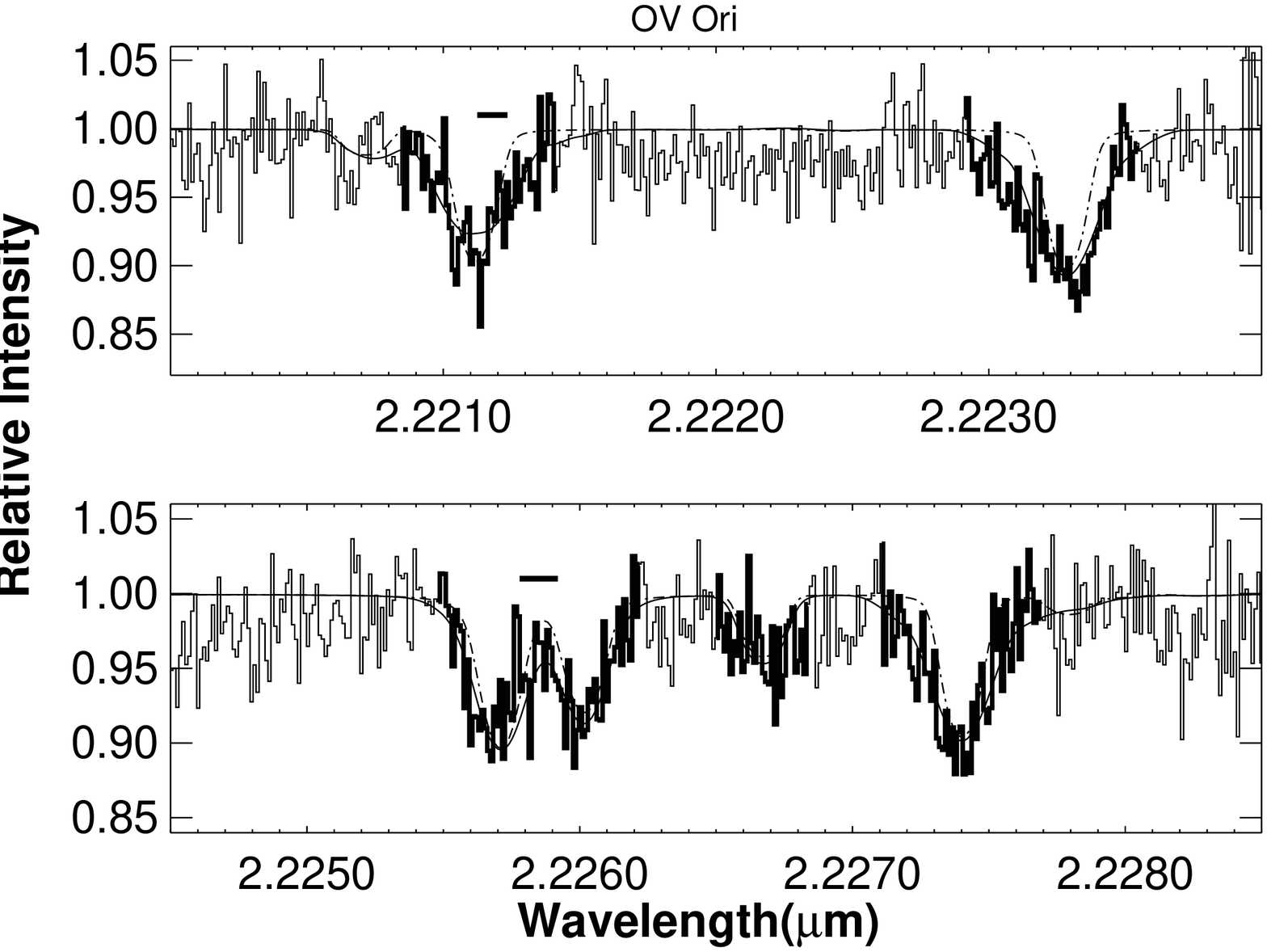}
   \caption {K-band infrared spectra of OV Ori. }
           \label{plot07}
   \end{center}
\end{figure}

\begin{figure}[ht]
  \begin{center}
    \includegraphics[scale=0.45]{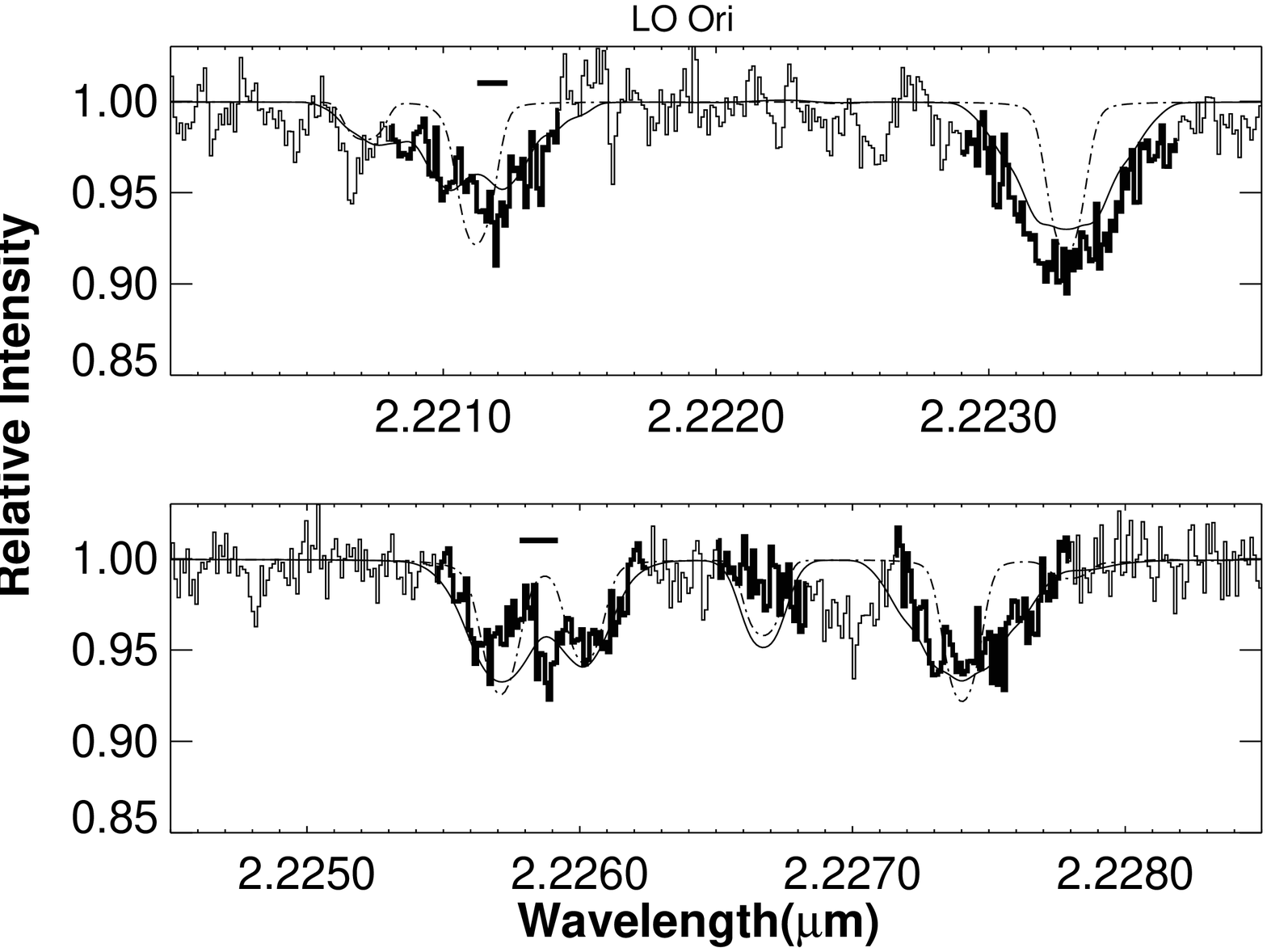}
   \caption {K-band infrared spectra of LO Ori. }
           \label{plot09}
   \end{center}
\end{figure}

\begin{figure}[ht]
  \begin{center}
    \includegraphics[scale=0.45]{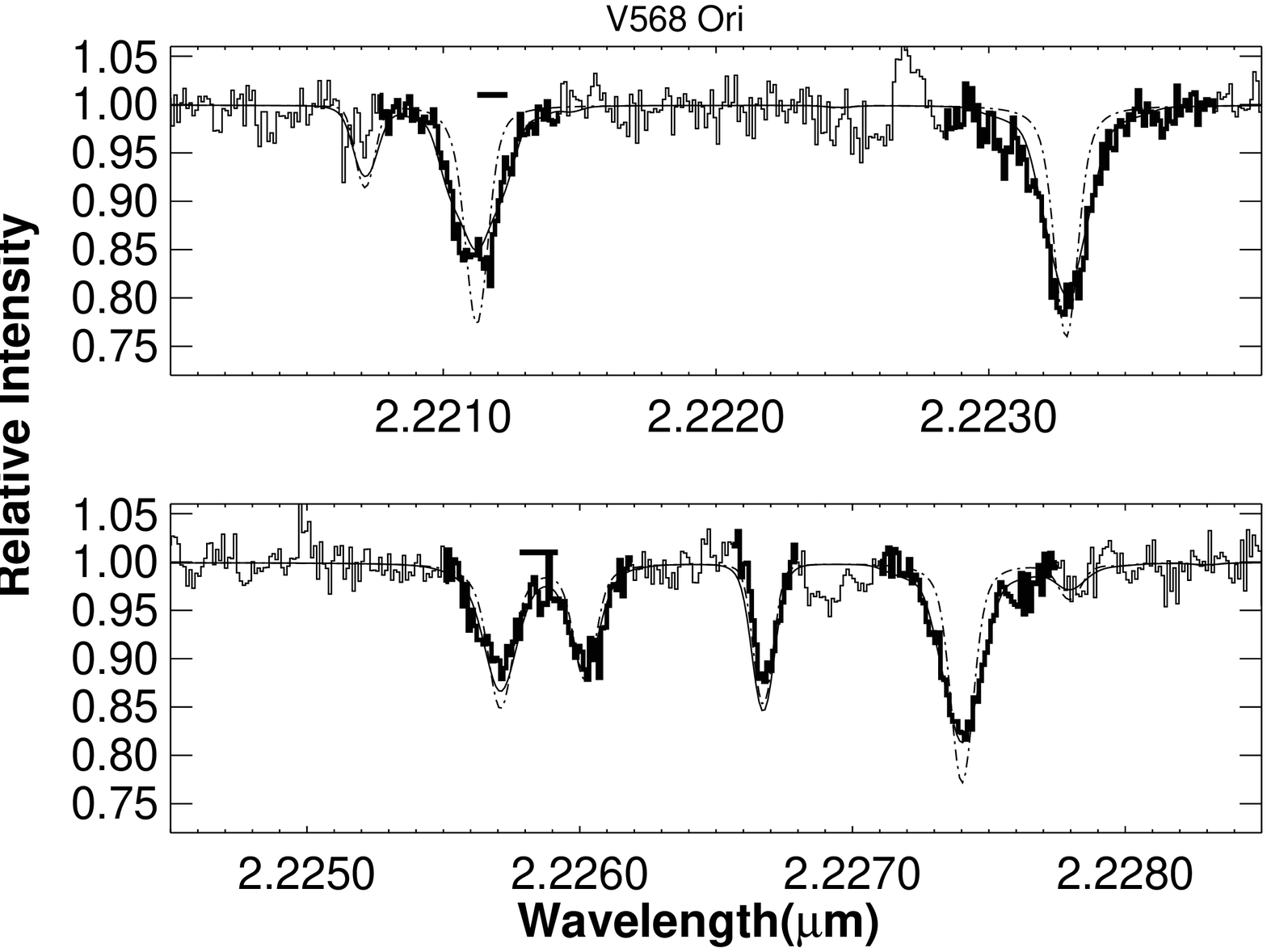}
   \caption {K-band infrared spectra of V568 Ori. }
           \label{plot10}
   \end{center}
\end{figure}

\begin{figure}[ht]
  \begin{center}
    \includegraphics[scale=0.45]{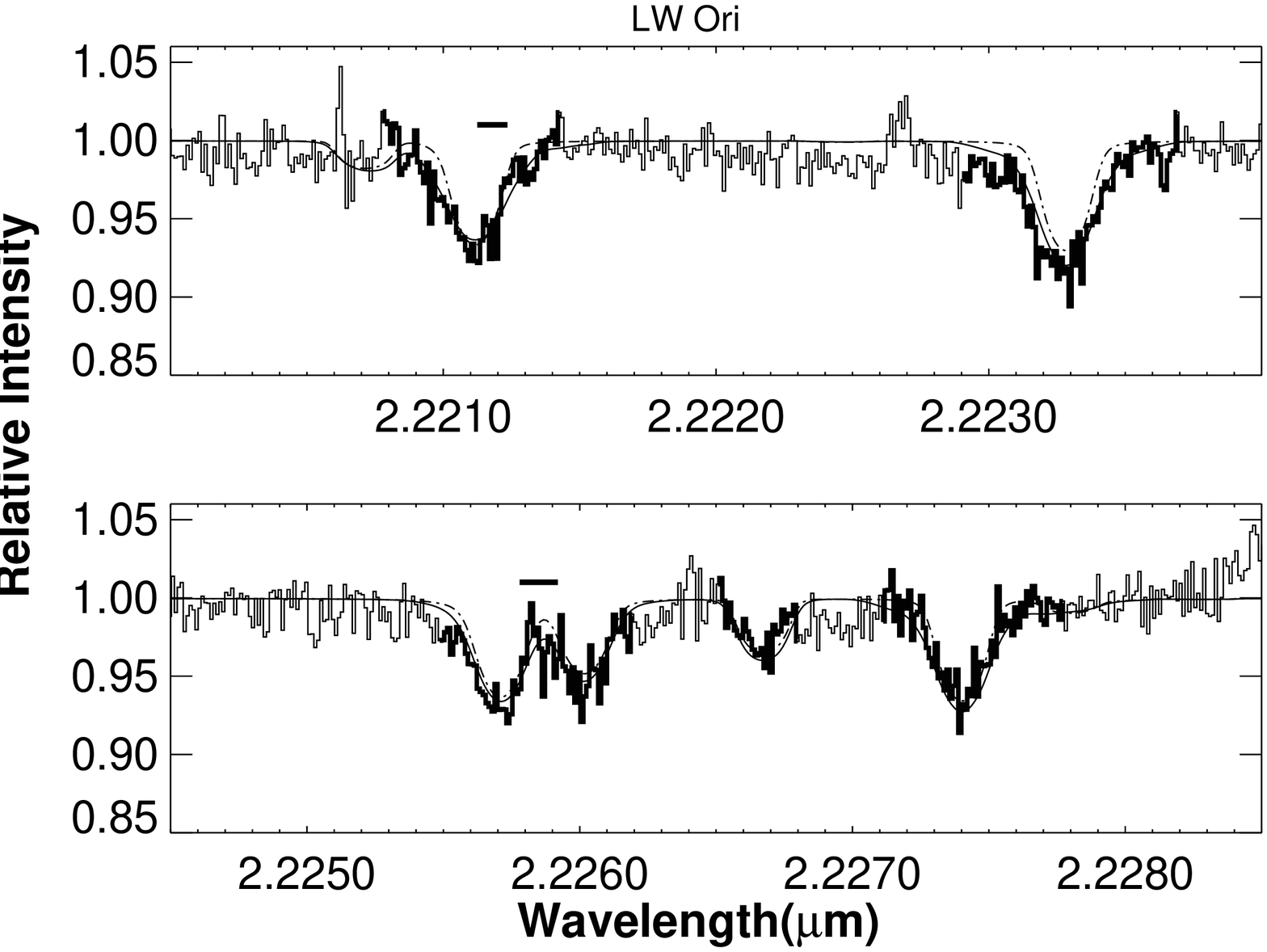}
   \caption {K-band infrared spectra of LW Ori. }
           \label{plot11}
   \end{center}
\end{figure}

\begin{figure}[ht]
  \begin{center}
    \includegraphics[scale=0.45]{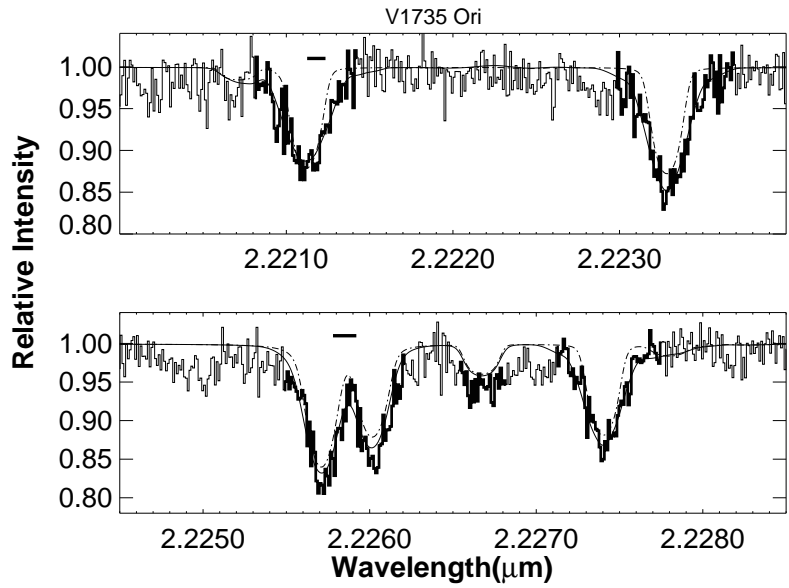}
   \caption {K-band infrared spectra of V1735 Ori. }
           \label{plot12}
   \end{center}
\end{figure}

\begin{figure}[ht]
  \begin{center}
    \includegraphics[scale=0.45]{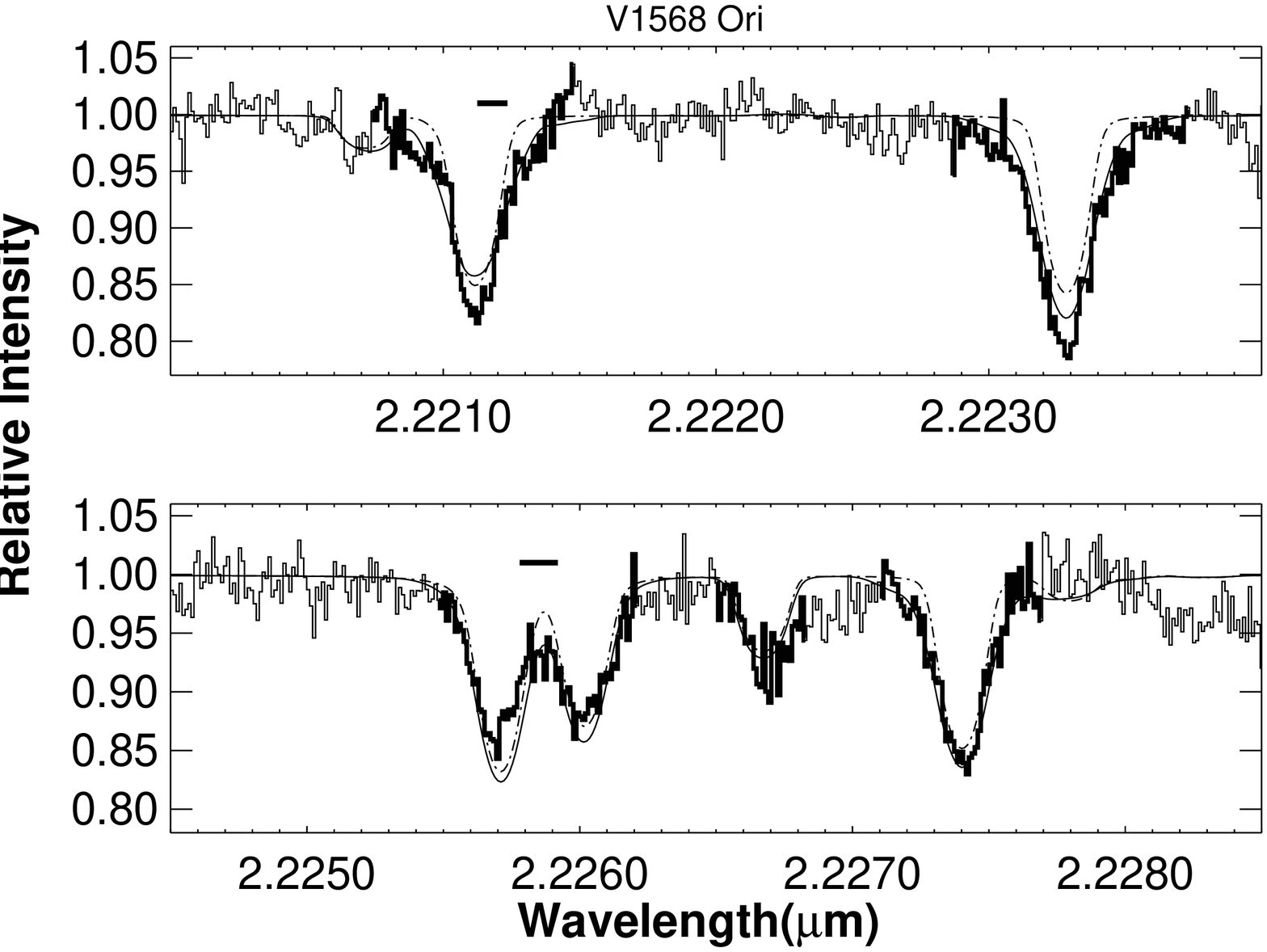}
   \caption {K-band infrared spectra of V1568 Ori. }
           \label{plot13}
   \end{center}
\end{figure}

\begin{figure}[ht]
  \begin{center}
    \includegraphics[scale=0.45]{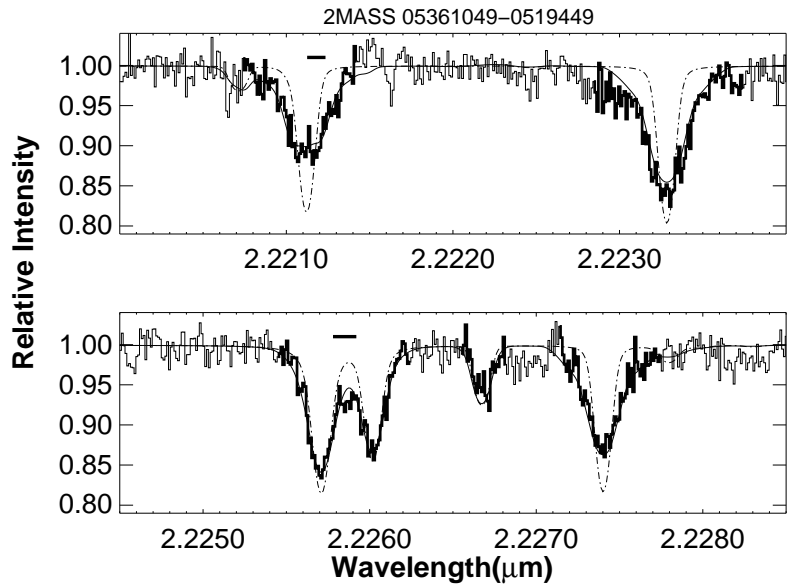}
   \caption {K-band infrared spectra of 2MASS 05361049-0519449. }
           \label{plot14}
   \end{center}
\end{figure}

\begin{figure}[ht]
  \begin{center}
    \includegraphics[scale=0.45]{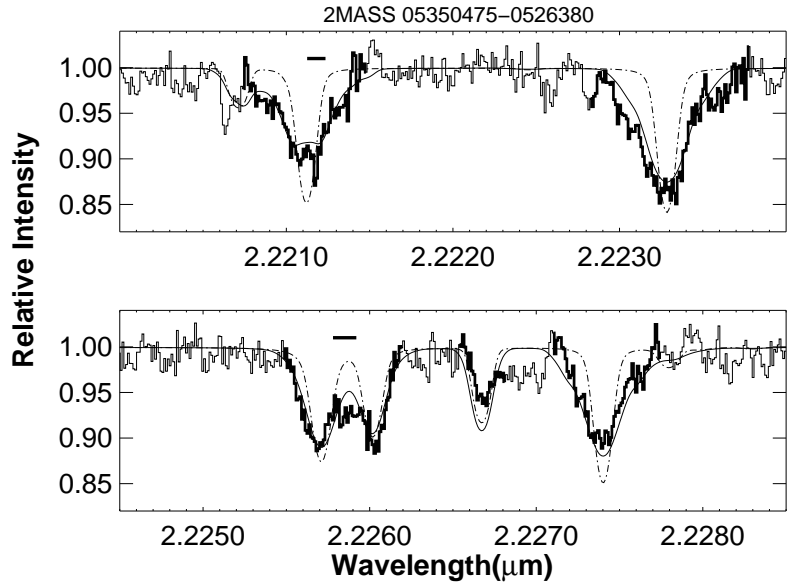}
   \caption {K-band infrared spectra of 2MASS 05350475-0526380. }
           \label{plot15}
   \end{center}
\end{figure}

\begin{figure}[ht]
  \begin{center}
    \includegraphics[scale=0.45]{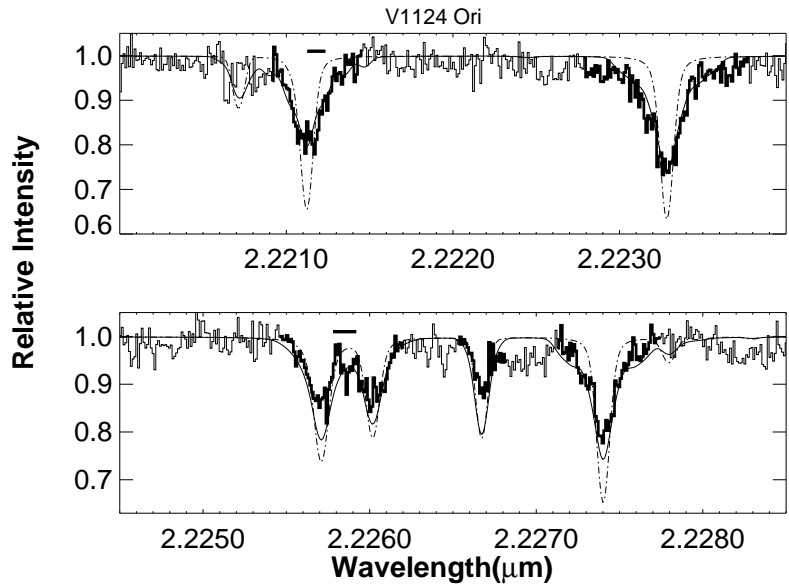}
   \caption {K-band infrared spectra of V1124 Ori. }
           \label{plot16}
   \end{center}
\end{figure}

\bibliographystyle{apj}
\bibliography{thesisref}

\end{document}